\documentclass[aip,amsmath,amssymb,reprint]{revtex4-2}
\usepackage{epsfig}
\usepackage{wrapfig}
\usepackage{bbm}
\usepackage[usenames]{color}
\usepackage{array}
\usepackage{times}

\usepackage{graphicx}
\usepackage{physics}
\usepackage{svg}
\usepackage{float}
\usepackage{multirow}
\usepackage{natbib,twoopt}
\usepackage[breaklinks=true]{hyperref}
\usepackage{xcolor}
\usepackage{amsmath,amsfonts,amssymb}
\usepackage{graphicx}
\usepackage{hyperref}
\usepackage{color}
\usepackage{stackengine}
\usepackage{subfigure}
\usepackage{verbatim}
\usepackage{stmaryrd} 

\newcommand{\rot}{\mathop{\rm rot}\nolimits}
\newcommand{\divv}{\mathop{\rm div}\nolimits}
\newcommand{\df}[2]{\frac{\partial #1}{\partial #2}}

\newcommand{\eps}{\varepsilon}
\newcommand{\om}{\omega}


\newcommand{\RNum}[1]{\uppercase\expandafter{\romannumeral #1\relax}}

\begin{document}

\title{Complex-valued Tellegen response}

\author{Fedor Nutskii}
\thanks{These authors have contributed equally to this work}
\affiliation{School of Physics and Engineering, ITMO University, Saint  Petersburg 197101, Russia}

\author{Eduardo Barredo-Alamilla}
\thanks{These authors have contributed equally to this work}
\affiliation{School of Physics and Engineering, ITMO University, Saint  Petersburg 197101, Russia}

\author{Maxim A. Gorlach}
\email{m.gorlach@metalab.ifmo.ru}
\affiliation{School of Physics and Engineering, ITMO University, Saint  Petersburg 197101, Russia}

\begin{abstract}
We consider a medium exhibiting non-reciprocal magneto-electric effect  known as Tellegen response captured by the equations of axion electrodynamics. Here, we investigate the implications of the complex-valued Tellegen response, discuss the conditions for its emergence and possible material realizations outlining a route to its experimental identification from the Stokes parameters of the reflected light.
\end{abstract}

\maketitle


Complex electromagnetic media known as metamaterials provide unique playground to obtain unconventional electromagnetic properties and craft the propagation of light flows. As the periodicity of metamaterials is much smaller than the wavelength, they are typically understood within the {\it effective medium} paradigm, which assigns them a set of effective material parameters~\cite{Simovski2009} obtained by the proper averaging of the microscopic fields~\cite{Silveirinha2007,Alu2011}. While this description is intrinsically approximate and suffers from many limitations including spatial dispersion effects~\cite{Belov2003}, it still could provide useful insights into the behavior of the composite structure.

A simple yet general framework for characterising such composite media is provided by the bianisotropic constitutive relations, which are typically presented in the form~\cite{Tretyakov}
\begin{equation}\label{mateq}
    \begin{gathered}
        \mathbf{D} = \tilde{\varepsilon}\,\mathbf{E} + \left(\tilde{\chi}+i\tilde{\kappa}\right)\,\mathbf{H}, \\
        \mathbf{B} = \left(\tilde{\chi}-i\tilde{\kappa}\right)^T \mathbf{E} + \mu\,\mathbf{H},\\
    \end{gathered}
\end{equation}
where $\tilde{\varepsilon}$ and $\mu$ are the permittivity and permeability, while $\tilde{\chi}$ and $\tilde{\kappa}$ quantify the bianisotropic response of the medium being responsible for the coupling between electric and magnetic degrees of freedom. Though in the general case all four material parameters are tensors, we focus on the scalar case below. If losses are absent, these material parameters are real~\cite{Tretyakov}.

Since electric and magnetic fields transform differently under inversion, the bianisotropic response could only occur if the inversion symmetry of the medium is broken. This appears to be sufficient to enable chirality $\tilde{\kappa}$ which is ubiquitous in nature in many organic molecules. On the contrary, nonreciprocity $\tilde{\chi}$ is much less common and requires not only broken inversion symmetry, but also broken time-reversal by static magnetic field, rotation or some other external influence.

The material with a scalar $\tilde{\chi}$ known as {\it Tellegen medium} is especially exotic. Though it was originally postulated long ago~\cite{tellegen1948gyrator}, the realistic proposals on its implementation in photonics were very limited~\cite{Tretyakov2003} until recently~\cite{Prudencio2023,Shaposhnikov2023, SafaeiJazi2024}, and first experimental realizations of the Tellegen-type media just started to appear~\cite{ShuangZhang2024,BaileZhang2025}. 

At the same time, such condensed matter structures as magnetoelectrics, multiferroics~\cite{Foner1963,Eerenstein2006, Pyatakov2012,Nenno2020} and topological insulators~\cite{Armitage2016,Sekine2021} do exhibit this kind of response, though the magnitude of the Tellegen response in those systems is relatively small, typically below $10^{-3}-10^{-2}$.

The constitutive relations Eqs.~\eqref{mateq} can be recast as
\begin{equation}\label{mateq1}
    \begin{gathered}
        \mathbf{D} = \varepsilon\,\mathbf{E} + \left(\chi+i\kappa\right)\,\mathbf{B}, \\
        \mathbf{H} = -\left(\chi-i\kappa\right) \mathbf{E} + \mu^{-1}\,\mathbf{B},\\
    \end{gathered}
\end{equation}
where $\chi=\tilde{\chi}/\mu$, $\kappa=\tilde{\kappa}/\mu$ and $\varepsilon=\tilde{\varepsilon}-(\tilde{\chi}^2+\tilde{\kappa}^2)/\mu$. Below, we term $\chi$ coefficient {\it Tellegen response}.

In this form [Eq.~\eqref{mateq1}], Maxwell's equations in the medium with $\kappa=0$ are identical to the equations of {\it axion electrodynamics}~\cite{Wilczek1987,Nenno2020} which describes the electromagnetic field in the presence of hypothetical axions~\cite{Wilczek1978,Weinberg1978}:
\begin{align}
&\rot {\left(\mu^{-1}\,\bf B\right)}=\frac{1}{c}\,\df{(\eps {\bf E})}{t}+\frac{4\pi}{c}\,{\bf j}\notag\\
&\mspace{90mu}+\left[\nabla \chi\times{\bf E}\right]+\frac{1}{c}\,\df{\chi}{t}\,{\bf B}\:,\label{eq:Axion1}\\
&\divv \left({\eps\,\bf E}\right)=4\pi\rho-\left(\nabla \chi\cdot {\bf B}\right)\:,\label{eq:Axion2}\\
&\rot {\bf E}=-\frac{1}{c}\,\df{{\bf B}}{t}\:,\mspace{10mu}\divv {\bf B}=0\:,\label{eq:Axion3}
\end{align}
where typically $\partial\chi/\partial t=0$ in metamaterial or condensed matter context. While cosmic axions are still to be observed, this  parallel renders condensed matter and metamaterial systems an interesting platform for probing axion physics, while the same coefficient $\chi$ is often called {\it axion response.}

However, the analogy between the physics of cosmic axions and its metamaterial or condensed matter implementations is incomplete. While in the high-energy physics the axion field is real acquiring an imaginary correction only in the very exotic circumstances~\cite{Brandenberger2021Apr,Bernardo2022Sep}, $\chi$ in metamaterials or condensed matter could potentially be complex. 

This possibility is briefly acknowledged in the metamaterial literature~\cite{lindell1994electromagnetic} and is largely uncharted in condensed matter~\cite{Nenno2020,ahn2022theory}, where the attention is focused on the quantization of the axion field $\chi$ and its connection to the topological invariants~\cite{Qi2008,Essin2009,Vazifeh-Franz,Devescovi2024,BaileZhang2025}, rather than the breakdown of this regime, inevitable for the frequency-dependent and lossy $\chi$. Notably, $\text{Im}\,\chi$ is fundamentally different from the chirality $\kappa$ as it enters the constitutive relations differently and requires breaking of the time-reversal symmetry.

Assumption of the complex-valued Tellegen response has several immediate consequences. First, nonzero $\text{Im}\,\chi$ leads to the dissipation~\cite{lindell1994electromagnetic}. However, since spatially homogeneous $\chi$ is not manifested in the bulk of the medium, this raises a question on where the dissipation happens. Second, since Tellegen response is not manifested in the bulk, it can be viewed as a surface Hall conductivity $\sigma_{xy}$ as sometimes done in the literature~\cite{Grushin2012Aug,Gong2023}. This raises a question whether bulk $\text{Im}\,\chi$ is equivalent to the imaginary correction to the Hall conductivity~\cite{Grushin2012Aug,Steiner2017Jul,Mukherjee2017Aug, Mukherjee2018Jan,ahn2022theory} or has a deeper connection to the bulk properties. 

In this Article, we address these issues by analyzing the physics behind complex $\chi$. As a specific model featuring isotropic Tellegen response, we explore an antiferromagnetic multilayered structure~\cite{Shaposhnikov2023}. We demonstrate that complex-valued $\chi$ is attainable, reveal the limitations on its magnitude and discuss how one can retrieve both $\text{Re}\,\chi$ and $\text{Im}\,\chi$ experimentally distinguishing the latter term from the conventional bulk losses associated with the imaginary parts of the permittivity or permeability.

\begin{figure}[b]
\center{\includegraphics[width=1\linewidth]{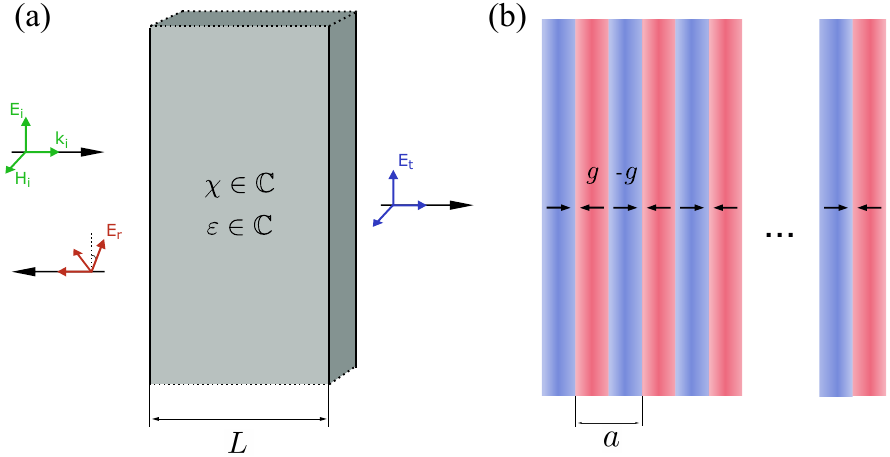}}
\caption{Probing effective Tellegen response by the incident plane wave. (a) Sketch of the structure and illustration of Kerr rotation for the reflected light. (b) Implementation of the structure based on gyrotropic layers with spatially alternating out-of-plane magnetization.}
\label{ris:image1}
\end{figure}

We start by examining a simple setup when a slab with the thickness $L$ featuring Tellegen response $\chi$, permittivity $\eps$ and permeability $\mu$ is illuminated by the linearly polarized plane wave at normal incidence incident from the medium with permittivity $\eps_0$ and permeability $\mu_0$ [Fig.~\ref{ris:image1}(a)]. In such case, the theory (see Supplementary Materials~\cite{Supplement}) predicts that the polarization of the transmitted light is unchanged, while the polarization of the reflected light experiences Kerr rotation by the angle
\begin{equation}\label{eq:KerrRotation}
    \tan \theta_r = \frac{2\sqrt{\dfrac{\varepsilon_0}{\mu_0}} \chi}{\dfrac{\varepsilon_0}{\mu_0} -\dfrac{\varepsilon}{\mu} - \chi^2}.
\end{equation}
%
Notably, this angle is independent of the slab thickness providing an easy access to quantify $\chi$.

If the medium is lossy, $\eps$, $\mu$ and, potentially, $\chi$ become complex. Therefore, the expression on the right-hand side of Eq.~\eqref{eq:KerrRotation} becomes complex. Physically, this means that the polarization of the reflected light becomes elliptic~\cite{Wu2016Dec, Perez-Garrido2023May} instead of linear, and the axes of the ellipse are rotated relative to the polarization plane of the incident light.

As a specific realization of the Tellegen medium we consider a multilayered structure composed of magneto-optical layers with antiparallel magnetization in the adjacent layers [Fig.~\ref{ris:image1}(b)]. Each layer is characterized by the gyrotropic permittivity tensor 
\begin{equation} \label{epsgyr}
    \hat{\varepsilon} = 
    \begin{pmatrix}
        \varepsilon & i g & 0\\
        -i g & \varepsilon & 0 \\
        0 & 0 & \varepsilon
    \end{pmatrix},
\end{equation}
where $g$ is a gyrotropy parameter. In this model, the Tellegen response emerges in the limit $\xi=a/\lambda\ll 1$ and reads~\cite{Shaposhnikov2023}
\begin{equation} \label{axeff}
    \chi_{\text{eff}} = \frac{\pi}{2}\, g\, \xi,
\end{equation}
while the effective permittivity of the structure reads
\begin{equation}\label{effectivepermittivity}
    \varepsilon_{\text{eff}} = \varepsilon + \frac{\pi^2}{12} g^2 \xi^2,
\end{equation}
and also incorporates second-order spatial dispersion contributions manifested at oblique incidence~\cite{Shaposhnikov2023} and not included in Eq.~\eqref{effectivepermittivity} for conceptual simplicity.

Complex-valued Tellegen response is attained when the gyrotropy $g$ becomes complex, $g=g'+ig''$. Physically, this captures the situation when orbital magnetic moments in the material precess experiencing the damping torque.

\begin{figure}[b!]
\center{\includegraphics[width=0.8\linewidth]{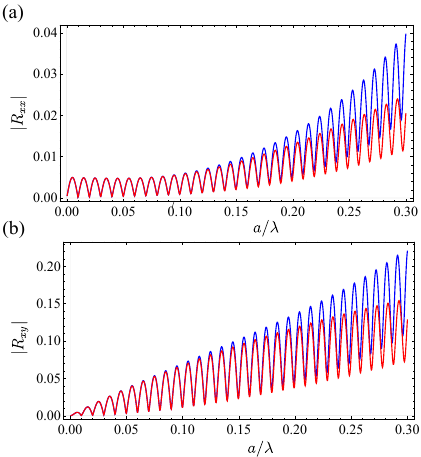}}
\caption{Co-polarized (a) and cross-polarized (b) reflection coefficients calculated from the effective medium description (red) and transfer matrix method (blue) with the parameters $g=0.5 + 0.01\,i$, $\eps = 1 + 0.01\,i , \varepsilon_0 = \mu_0 = \mu = \mu_\text{eff}  = 1$,  $\chi_\text{eff}$, and $\eps_{\text{eff}}$ are given by Eqs.~ (\ref{axeff})-(\ref{effectivepermittivity}), total number of periods $N=50$. 
}
\label{ris:image2}
\end{figure}

However, complex correction to gyrotropy $g''$ cannot arise on its own and is necessarily accompanied by the complex contribution $\eps''$ to the permittivity. This conclusion is readily derived from the passivity condition~-- the requirement that the medium can only dissipate energy, but does not provide gain (see Supplementary Materials, Sec.~S3~\cite{Supplement}). Using $e^{-i\om t}$ time convention for the fields, we obtain $\eps''>0$ and
\begin{equation} \label{gyrlim}
|g''| \leq \varepsilon''\:.
\end{equation}

To validate the accuracy of the effective medium description of the multilayered structure with complex $g$, we compare the reflection coefficients for the structure calculated via rigorous transfer matrix method [see Supplementary Materials~\cite{Supplement}, Sec.~S4] with the respective effective medium predictions.  We study the structure consisting of $50$ unit cells, each is composed  of two gyrotropic layers of equal thickness $a/2$ and opposite magnetizations $\pm g_0$, $g_0 = 0.5 + 0.01 \, i$.

The results shown in Fig.~\ref{ris:image2} illustrate the behavior of co-polarized and cross-polarized reflection coefficients as a function of the period-to-wavelength ratio $\xi$ in the subwavelength range from 0 to 0.3. Practically, this ratio can be varied continuously by changing the frequency of the incident wave. In the limit $\xi \ll 1$,  the rigorous solution (blue lines) closely matches the predictions of the effective medium model (red lines). Deviations increase with $\xi$ becoming noticeable when $\xi$ approaches $0.15$. A similar trend is observed for the transmission spectra as further elaborated in the Supplementary Materials~\cite{Supplement}, Sec.~S4.

\begin{figure}[b!]
\center{\includegraphics[width=0.5\linewidth]{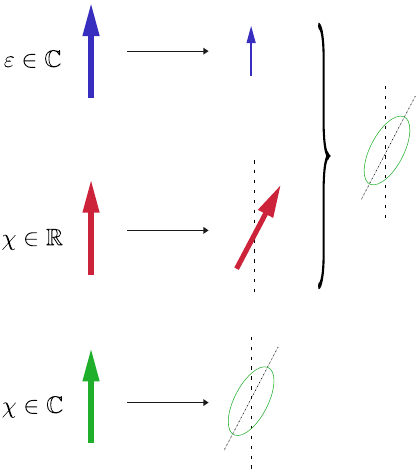}}
\caption{Schematic showing the key factors affecting the polarization of light reflected from the surface of the medium with complex Tellegen response.}
\label{ris:image3}
\end{figure}

Thus, the stack of magneto-optical layers with complex gyrotropy matches the effective medium model in the limit $\xi\ll 1$, realizes complex Tellegen response and hence provides a convenient model to study the associated physics. Using this system, we now examine how complex Tellegen response can be detected experimentally and distinguished from the other linear responses. Fig.~\ref{ris:image3} depicts schematically the key ingredients affecting the reflection of the linearly polarized light.

Imaginary part of the medium permittivity $\eps''$ leads to the bulk losses and hence decreases the amplitude of both transmitted and reflected waves. However, the polarization of reflected light still remains linear in the same plane. 

Real Tellegen response $\chi'$ causes the rotation of polarization plane. Hence, the combination of $\chi'$ and $\eps''$ results generally in elliptical polarization. 

Finally, complex Tellegen response $\chi''$ both rotates the polarization plane and transformes the linear polarization into the elliptical one.

This simple reasoning raises a question on whether it is possible to experimentally distinguish a complex Tellegen response from the combination of the two conventional responses~-- complex permittivity $\eps$ and purely real $\chi$.
%

To address that, we characterize the reflected light by the set of Stokes parameters
\begin{equation}\label{eq:stokes}
    \begin{gathered}
        I_R = |E_x|^2 + |E_y|^2, \\
        Q = |E_x|^2 - |E_y|^2, \\
        U = 2 \Re{E_x E_y^*}, \\
        V = -2 \Im{E_x E_y^*},
    \end{gathered}
\end{equation}
capturing the intensity of the reflected light ($I_R$) and its polarization state ($Q$, $U$, $V$). These characteristics related via $I_R^2=Q^2+U^2+V^2$ which defines the Poincar{\'e} sphere for the reflected light. In addition, we assume that the intensity of the transmitted light $I_T$ is also known.

Four independent characteristics of the reflected and transmitted light suggest that the four relevant parameters~-- namely, $\eps'$, $\eps''$, $\chi'$ and $\chi''$~-- can be retrieved independently. In Sec.~S5 of the Supplementary Materials~\cite{Supplement} we perform such a retrieval analytically assuming weak Tellegen response $\chi$, i.e. neglecting terms $\propto\chi^2$, and assuming low losses, $\varepsilon''\ll|\varepsilon'|$. Real and imaginary parts of the permittivity are recovered from the intensity of reflected and transmitted light, while the Tellegen response is extracted from the Stokes parameters $U$ and $V$:
%
\begin{align}
     & \chi' = - \frac{(1-\varepsilon')\, U + \varepsilon''\, V }{4\, I_R}, \\
    & \chi''  = -\frac{(1-\varepsilon') V - \varepsilon''\, U}{4\, I_R}\:.
\end{align}
%
%
\begin{figure}[b!]
\center{\includegraphics[width=1\linewidth]{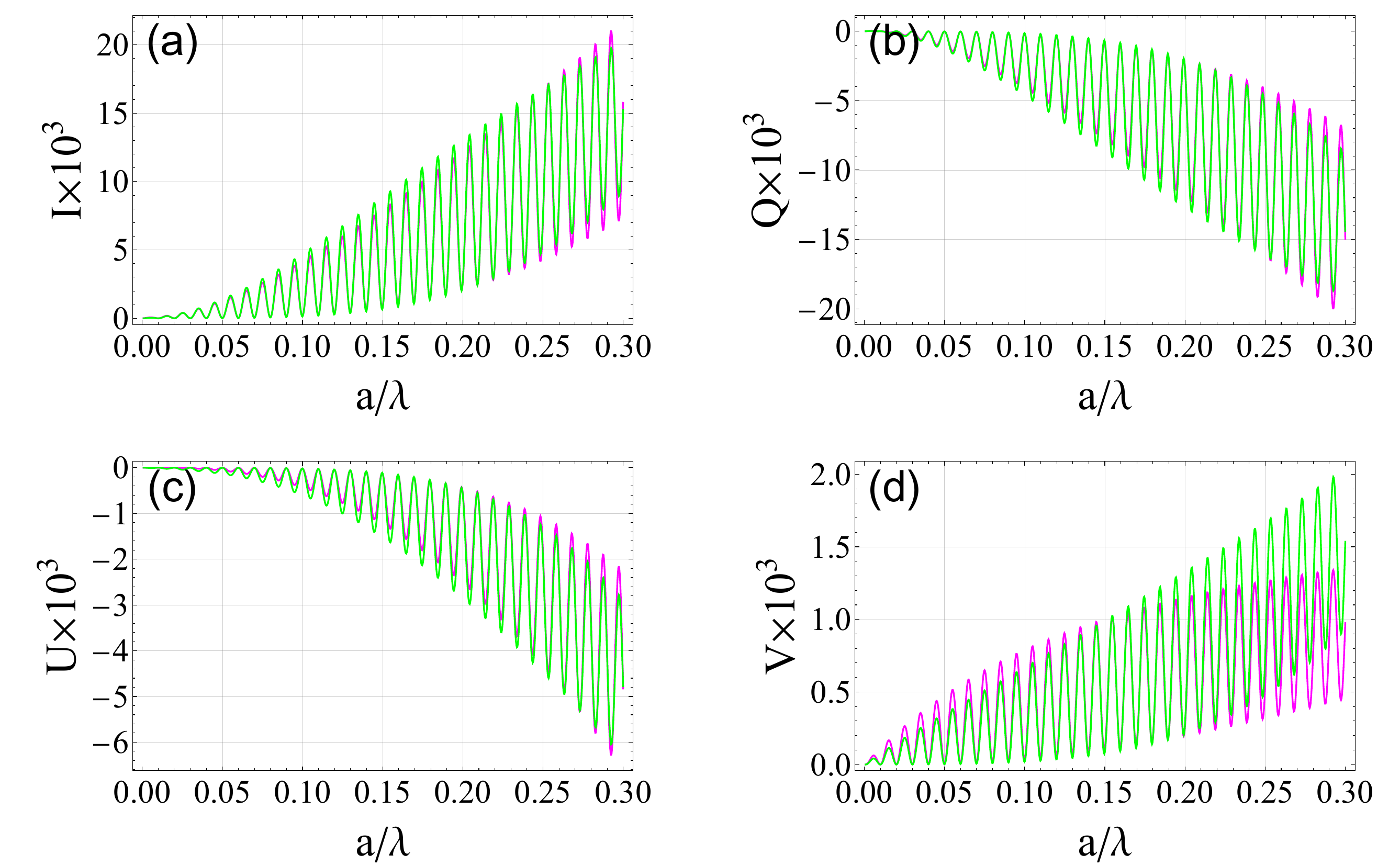}}
\caption{Stokes parameters for a wave reflected from the multilayered structure featuring complex Tellegen response. Green lines: results for a medium with complex  $\chi$ and complex $\eps$.  Magenta lines: Attempt to fit the data assuming a real $\chi$ and complex permittivity $\eps$. Green curves: $g_0 = 0.5 + 0.01 i$, $\eps = 1 +0.01 i$. Magenta curves: $g_0 =0.5$, $\eps = 1 + 0.015 i$. A total of 50 unit cells were used in the simulation.}
\label{ris:image4}
\end{figure}

To illustrate the retrieval numerically, we compute the Stokes parameters for light reflected from a medium with complex permittivity $\eps_{\text{eff}}$ and complex Tellegen response. In Fig.~\ref{ris:image4}, the green curves depict the Stokes parameters computed for the medium consisting of 50 pairs of oppositely magnetized layers, each with the thickness $a/2$, gyrotropy amplitude $g_0 = 0.5 + 0.01 i$ and the permittivity $\varepsilon= 1+ 0.01 i$. For comparison, we attempt to fit the same data using a simplified model which assumes purely real Tellegen response, and associates all losses with the complex permittivity $\eps = 1+0.015 i$ as shown by magenta curves in Fig.~\ref{ris:image4}.


We observe that not all Stokes parameters are accurately fitted, indicating the relevance of $\chi''$ to capture the optical properties of the medium.

In the analyzed system, complex Tellegen response  $\chi''$ requires complex gyrotropy $g''$, which necessarily leads to the nonzero imaginary part of the permittivity $\eps''$. Notably, such situation is generic, and complex Tellegen response $\chi$ is always accompanied by the complex permittivity and permeability. To demonstrate that, we analyze the energy balance in the medium with complex material parameters (see Supplementary Materials~\cite{Supplement}, Sec.~S3) and require that for any field configuration the energy is dissipated. This yields the inequality
\begin{equation}\label{eq:ChiConst}
|\chi''| \leq \frac{\sqrt{\varepsilon'' \mu''}}{|\mu|}.
\end{equation}
Equation~\eqref{eq:ChiConst} demonstrates that the imaginary contribution to the Tellegen response ($\chi'' \neq 0$) necessarily requires nonzero imaginary parts of the two bulk parameters~-- permittivity ($\varepsilon''>0$) and permeability ($\mu''>0$). Interestingly, the real part of $\chi$ remains unconstrained, decoupling its behavior from the dissipation-driven dynamics of the imaginary part and allowing to obtain strong Tellegen response, as has been recently demonstrated experimentally~\cite{ShuangZhang2024} and justified theoretically~\cite{Seidov2024}.

On a side note, the considered multilayered structure does not have an effective permeability, but features second-order spatial dispersion effects instead~\cite{Shaposhnikov2023}. However, the general conclusion stays the same: $\chi''$ is necessarily accompanied by the bulk losses.

In summary, we have demonstrated the emergence of complex Tellegen response in a metamaterial composed of layers with complex gyrotropy. The key ingredients for obtaining complex $\chi$ in our case are the damping of magnetization precession described by the complex gyrotropy $g$ and antiferromagnetic pattern of magnetization. Qualitatively similar physics may occur in condensed matter systems, where damping of anti-scalar magnons enables imaginary correction to the axion response.


Our results suggest that complex $\chi$ cannot arise on its own and is necessarily accompanied by the imaginary corrections to  other material parameters, such as permittivity and permeability. These additional sources of loss potentially complicate the  observation of complex $\chi$. Still, measuring the polarization of the reflected light, one could distinguish complex Tellegen response from the other material responses.

The connection between  $\chi''$ and imaginary corrections to the bulk material parameters suggests that complex Tellegen response should rather be viewed as a bulk property of the medium. In the same logic, losses associated with the complex $\chi$ happen not in the thin surface layer, but in the bulk of the structure. This conclusion is evident not only from our model of a multilayered structure with the complex gyrotropy, but also from the expression of the time-averaged energy flow
\begin{equation}
   \mathbf{S}=\frac{c}{8\pi}\,\text{Re}\left[\mu^{-1}\,\mathbf{E}^*\times\mathbf{B}\right]-\frac{c}{8\pi}\,\text{Re}\left[\chi\,\mathbf{E}^*\times\mathbf{E}\right]\:, 
\end{equation}
which incorporates an additional term $\propto\chi$, avoids the discontinuity of the energy flow at the boundary of the medium, but enables the dissipation in the bulk.

Our findings thus clarify the issues related to complex Tellegen response and provide a theoretical foundation for future experimental investigations of Tellegen metamaterials and axion insulators.



\section*{Acknowledgments}

Theoretical models were supported by the Russian Science Foundation (Grant No.~23-72-10026). Numerical simulations were supported by Priority 2030 Federal Academic Leadership Program. E.B.-A. acknowledges partial support from CONAHCYT (México) doctoral scholarship. M.A.G. acknowledges partial support from the Foundation for the Advancement of Theoretical Physics and Mathematics ``Basis''.

\section*{References}
\bibliography{Complex_axion}

\end{document}


\title{Supplementary Materials: \\ 
Complex-valued Tellegen response}

\author{Fedor Nutskii}
\affiliation{School of Physics and Engineering, ITMO University, Saint  Petersburg 197101, Russia}

\author{Eduardo Barredo-Alamilla}
\affiliation{School of Physics and Engineering, ITMO University, Saint  Petersburg 197101, Russia}

\author{Maxim A. Gorlach}
\email{m.gorlach@metalab.ifmo.ru}
\affiliation{School of Physics and Engineering, ITMO University, Saint  Petersburg 197101, Russia}


\maketitle
\widetext

\tableofcontents

\section{Reflection and transmission of a plane wave at complex axion interface}

\begin{figure}[h!]
  \centering
\includegraphics[width=0.70\textwidth]{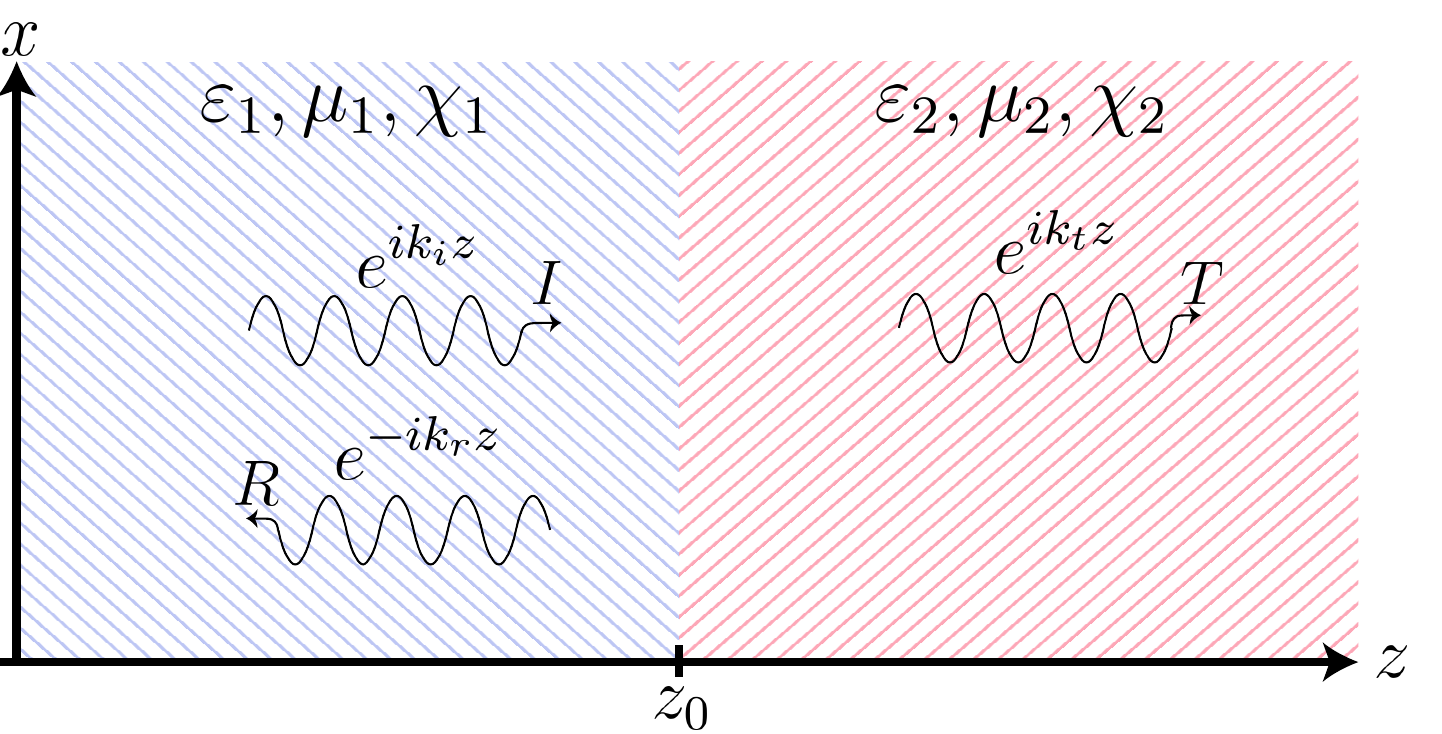}
  \caption{Reflection and transmission of a plane wave at the boundary separating two  media with the different axion responses $\chi_1$ and $\chi_2$.} 
  \label{fig:axiontimeboundary}
\end{figure}

We consider two semi-infinite media with complex axion response, characterized by the material parameters $\chi_1$, $\varepsilon_1$, $\mu_1$ for $z<z_0$ and $\chi_2$, $\varepsilon_2$, $\mu_2$  for $z>z_0$. Our goal is to derive the reflection and transmission coefficients for a plane wave incident at the interface $z=z_0$.

The boundary conditions derived from the equations of axion electrodynamics read:
\begin{align}
   [\mathbf{D}]_n & = \chi \mathbf{B}\cdot \hat{\mathbf{n}}|_{z=z_0}, & [\mathbf{H}]_\parallel & = -\chi \mathbf{E}\times \hat{\mathbf{n}}|_{z=z_0}, \label{AxionBC1} \\
   [\mathbf{B}]_n & = 0, & [\mathbf{E}]_\parallel & = 0, \label{AxionBC2} 
\end{align}
where $\chi = \chi_1-\chi_2$ and $[\mathbf{D}]_n\equiv D_{2n}-D_{1n}$. For normal incidence, assuming that the wave propagates along $\hat{\mathbf{z}}$, the reflection and transmission matrices are computed using the transfer matrix method. The plane wave solutions are:
\begin{align}
    \mathbf{E}^1 = & \mathbf{E}_{i}(z,t) = \mathbf{E}_i(k_i,\omega) e^{i k_i z} e^{-i \omega t}, \\
    \mathbf{E}^2 = & \mathbf{E}_{t}(z,t) + \mathbf{E}_{r}(z,t)= \mathbf{E}_t(k_t,\omega) e^{i k_t z} e^{-i \omega t} + \mathbf{E}_r(k_r,\omega) e^{-i k_r z} e^{-i \omega t},
\end{align}
where the subscripts $i$, $r$ and $t$ indicate incident, reflected and transmitted waves. From the bulk Maxwell's equations the propagation constants of the waves are found to be $k_i = k_0 \sqrt{\mu_1\varepsilon_1}$ and $k_r = k_t = k_0 \sqrt{\mu_2\varepsilon_2}$, where $k_0=\omega/c$. Using the boundary conditions for the tangential components in Eqs.~\ref{AxionBC1}, the amplitudes of the incident, transmitted, and reflected waves are related as
\begin{align}
    \vc{E}_t  & = \vc{E}_i + \vc{E}_r,\label{AxionBoundary1}\\
    \sqrt{\frac{\varepsilon_2}{\mu_2}} \vc{E}_t & =  \chi \hat{\mathbf{n}} \times (\mathbf{E}_i+\vc{E}_t) + \sqrt{\frac{\varepsilon_1}{\mu_1}}(\vc{E}_i -\vc{E}_r).\label{AxionBoundary2}
\end{align}
Therefore, we can construct the axion-boundary transfer matrix directly from Eqs. (\ref{AxionBoundary1}), (\ref{AxionBoundary2})  choosing as an appropriate basis the vectors:
\begin{equation}
    \begin{pmatrix}
        \vphantom{\sqrt{\dfrac{\varepsilon_1}{\mu_1}}} \mathbf{E}_t  \\ \sqrt{\dfrac{\varepsilon_2}{\mu_2}} \mathbf{E}_t   
    \end{pmatrix} =  \begin{pmatrix}
        \vphantom{\sqrt{\dfrac{\varepsilon_1}{\mu_1}}}M_{11} & M_{12}  \\
        M_{21} & \vphantom{\sqrt{\dfrac{\varepsilon_1}{\mu_1}}}M_{22}
    \end{pmatrix}
     \begin{pmatrix}
        \vphantom{\sqrt{\dfrac{\varepsilon_1}{\mu_1}}} \mathbf{E}_i +\mathbf{E}_r\\ \sqrt{\dfrac{\varepsilon_1}{\mu_1}} \Big( \mathbf{E}_i -\mathbf{E}_r\Big)
    \end{pmatrix}
\label{transfermatrixeqaxion},\end{equation}
where $M_{11}=\mathbb{1}$, $M_{12}=0$, $M_{22}=\mathbb{1}$, and

\begin{equation}
    M_{21} = \chi \begin{pmatrix}
        \vphantom{\sqrt{\frac{\varepsilon_1}{\mu_1}}}0 & -1  \\
        1 & \vphantom{\sqrt{\frac{\varepsilon_1}{\mu_1}}}\hphantom{-}0
    \end{pmatrix}.
\end{equation}
This choice of basis simplifies the boundary conditions, enabling a straightforward determination of the reflection and transmission matrices: 
\begin{align}
    \hat{R} & = -\left[\sqrt{\frac{\varepsilon_2}{\mu_2}} M_{11} - \sqrt{\frac{\varepsilon_1\varepsilon_2}{\mu_1\mu_2}} M_{12} - M_{21} + \sqrt{\frac{\varepsilon_1}{\mu_1}} M_{22}\right]^{-1} \left[\sqrt{\frac{\varepsilon_2}{\mu_2}} M_{11} + \sqrt{\frac{\varepsilon_1\varepsilon_2}{\mu_1\mu_2}} M_{12} - M_{21} - \sqrt{\frac{\varepsilon_1}{\mu_1}} M_{22}\right], \\
    \hat{T} & = \left[M_{11} +\sqrt{\frac{\varepsilon_1}{\mu_1}}M_{12} + \left( M_{11}-\sqrt{\frac{\varepsilon_1}{\mu_1}} M_{12} \right)\hat{R}\right],
\end{align}
that describe the transmitted and reflected fields $\mathbf{E}_t = \hat{T}\;\mathbf{E}_i$ and $\mathbf{E}_r = \hat{R} \; \mathbf{E}_i$. The full form of these matrices is

\begin{align}
    \hat{T} & =  \frac{2\sqrt{\dfrac{\varepsilon_1}{\mu_1}}}{\left(\sqrt{\dfrac{\varepsilon_1}{\mu_1}}+\sqrt{\dfrac{\varepsilon_2}{\mu_2}}\right)^2+\chi^2} 
    \begin{pmatrix}[2]
 \sqrt{\dfrac{\varepsilon_2}{\mu_2}} +\sqrt{\dfrac{\varepsilon_1}{\mu_1}} & -\chi  \\[10pt]
 \chi & \sqrt{\dfrac{\varepsilon_2}{\mu_2}} +\sqrt{\dfrac{\varepsilon_1}{\mu_1}} \\
    \end{pmatrix}, \\[5pt]
    \hat{R} & =  \frac{1}{\left(\sqrt{\dfrac{\varepsilon_1}{\mu_1}}+\sqrt{\dfrac{\varepsilon_2}{\mu_2}}\right)^2+\chi^2} 
    \begin{pmatrix}[3]
  \dfrac{\varepsilon_1}{\mu_1}-\dfrac{\varepsilon_2}{\mu_2} - \chi^2 & -2\chi\sqrt{\dfrac{\varepsilon_1}{\mu_1}}  \\
 2\chi\sqrt{\dfrac{\varepsilon_1}{\mu_1}} & \dfrac{\varepsilon_1}{\mu_1}-\dfrac{\varepsilon_2}{\mu_2}  - \chi^2 \\
    \end{pmatrix}.
\end{align}
Using the Poynting vector
\begin{equation}
    \vc{S} = \frac{c}{8\pi} \text{Re} (\vc{E} \times \vc{H}^*),
\end{equation}
we compute the intensities of reflected and transmitted waves. Note that the $\vc{H}$ field in the axion medium is $\vc{H} = \mu^{-1} \vc{B} - \chi \vc{E}$ and therefore

\begin{equation}
    \vc{S} = \frac{c}{8\pi} \frac{1}{k_0}\text{Re}\left[\frac{\vc{k}}{\mu} |\vc{E}|^2 \right] - \frac{c}{8\pi} \text{Re}\left[\vphantom{\frac{\vc{k}}{\mu}} \chi^* \vc{E}\times \vc{E}^*\right].
\end{equation} 
It should be stressed that $\vc{E}\times\vc{E}^*$ could be nonzero for elliptical or circular polarization. Since $\vc{E}\times\vc{E}^*$ is purely imaginary, an additional term in the energy flow arises only for the complex $\chi$.


Naively, one may assume that the axion term $\chi$ is complex, while the permittivity and permeability are purely real. Below, we demonstrate that this assumptions leads to the inconsistencies. Specifically, the energy flow for the incident, reflected, and transmitted waves is evaluated as.
\begin{align}
    S_{i} & = \mathbf{S}_i \cdot \mathbf{\hat{z}} = \frac{c}{8\pi} \sqrt{\frac{\varepsilon_1}{\mu_1}} |\mathbf{E}_i|^2,\\
    S_{t} & = \mathbf{S}_t \cdot \mathbf{\hat{z}} = \frac{c}{8\pi} \sqrt{\frac{\varepsilon_2}{\mu_2}} |\mathbf{E}_t|^2 - \frac{c}{8\pi} \text{Re}\left[\vphantom{\frac{\vc{k}}{\mu}} \chi_2^* \vc{E}_t\times \vc{E}_t^*\right], \\
     S_{r} & = \mathbf{S}_r \cdot \mathbf{\hat{z}} = \frac{c}{8\pi} \sqrt{\frac{\varepsilon_1}{\mu_1}} |\mathbf{E}_r|^2 - \frac{c}{8\pi} \text{Re}\left[\vphantom{\frac{\vc{k}}{\mu}} \chi_1^* \vc{E}_r\times \vc{E}_r^*\right],
\end{align}
where the axion response is assumed complex and decomposed into real and imaginary parts as $\chi_{\alpha} = \chi_{\alpha,r} + i \chi_{\alpha,i}$, for $\alpha=1,2$. In addition, the polarization of the incident light is linear, $\vc{E}^*_i = \vc{E}_i$. Intensity transmission and reflection coefficients are:
\begin{align}
    T & = \frac{S_t}{S_i}, & R & = \frac{S_r}{S_i}.
\end{align}
To compute these coefficients explicitly, we deduce

\begin{align}
    \vc{E}_r\times \vc{E}_r^* & =   2 i \;\text{Im}\Big[ r_{12} r_{11}^* \Big] |\vc{E}_i|^2, \\
     \vc{E}_t\times \vc{E}_t^* & =  2 i \; \text{Im} \Big[ t_{12} t_{11}^* \Big] |\vc{E}_i|^2,
\end{align}
and also
\begin{align}
    \vc{E}_r \cdot \vc{E}_r^* & = \Big(|r_{11}|^2 + |r_{12}|^2 \Big) |\vc{E}_i|^2, \\
    \vc{E}_t \cdot \vc{E}_t^* & =\Big(|t_{11}|^2 + |t_{12}|^2\Big) |\vc{E}_i|^2.
\end{align}
Therefore, in terms of the amplitude reflection and transmission coefficients the result reads:
\begin{align}
 R & =   \Big(|r_{11}|^2 + |r_{12}|^2\Big) |\vc{E}_i|^2 + 2  \;\sqrt{\frac{\mu_1}{\varepsilon_1 }}  \chi_{1,i} \;  \text{Im} \Big[ r_{12} r_{11}^* \Big] |\vc{E}_i|^2, \label{rcoefficient}\\
 T & =  \frac{\sqrt{\dfrac{\varepsilon_2}{\mu_2}}}{\sqrt{\dfrac{\varepsilon_1}{\mu_1}}} \Big(|t_{11}|^2 + |t_{12}|^2\Big) |\vc{E}_i|^2 +  2  \; \sqrt{\frac{\mu_1}{\varepsilon_1}}  \chi_{2,i} \; \text{Im}  \Big[ t_{12} t_{11}^* \Big] |\vc{E}_i|^2, \label{tcoefficient}
\end{align}
where $\chi_1 = \chi _{1,r} + i\chi_{1,i}$ and $\chi_2 = \chi _{2,r} + i\chi_{2,i}$. Now, we just substitute the values of $r_{11}, r_{12}, t_{11}, t_{12}$, obtaining

\begin{align}
    R & = \frac{1}{|f(\chi)|^2} \left[ \left( \frac{\varepsilon_1}{\mu_1} - \frac{\varepsilon_2}{\mu_2}\right)^2 - \left( \frac{\varepsilon_1}{\mu_1} - \frac{\varepsilon_2}{\mu_2}\right) \bigg(\chi^2 + (\chi^*)^2\bigg) + |\chi|^4 + 4\frac{\varepsilon_1}{\mu_1} |\chi|^2 + 4\chi_i \chi_{i,1} \left( \frac{\varepsilon_2}{\mu_2} - \frac{\varepsilon_1}{\mu_1} - |\chi|^2 \right) \right]. \\
     T & = \frac{1}{|f(\chi)|^2} \left[ 4\sqrt{\frac{\varepsilon_1 \varepsilon_2}{\mu_1 \mu_2}} \left\{ \left( \frac{\varepsilon_1}{\mu_1} - \frac{\varepsilon_2}{\mu_2}\right)^2 + |\chi|^2 \right\} + 8 \sqrt{\frac{\varepsilon_1}{\mu_1}} \chi_i \chi_{2,i} \left(\sqrt{\frac{\varepsilon_1}{\mu_1}} + \sqrt{\frac{\varepsilon_2}{\mu_2}} \right)\right],
\end{align}
where
\begin{align}
    |f(\chi)|^2 = \left|  \left(\sqrt{\frac{\varepsilon_1}{\mu_1}} + \sqrt{\frac{\varepsilon_2}{\mu_2}} \right)^2 + \chi^2 \right|^2,
\end{align}
and $\chi_i = \chi_{1,i}-\chi_{2,i}$.

Interestingly, the energy is generally not conserved and 
\begin{equation}
    T + R = 1 + \frac{4 \chi_i \chi_{1,i}}{|f(\chi)|^2} \left\{ \left(\sqrt{\frac{\varepsilon_1}{\mu_1}} + \sqrt{\frac{\varepsilon_2}{\mu_2}} \right)^2 - |\chi|^2\right\}. 
\label{conservationcomplexinterface}\end{equation}

To illustrate the consequences, we analyze several representative cases. We assume strong Tellegen response $|\chi| > \left|\sqrt{\frac{\varepsilon_1}{\mu_1}} + \sqrt{\frac{\varepsilon_2}{\mu_2}}\right|$, with $|\chi| = \sqrt{\chi_r^2 + \chi_i^2}$, where $\chi_r = \chi_{1,r}-\chi_{2,r}$.
\begin{itemize}
\item {\bf Case A:}  If $\chi_{1,i} = 0$, then $T+R = 1$ and the energy is conserved.
\item {\bf Case B:}  If $\chi_{1,i} > 0$, and $\chi_{2,i} = 0$, then 

\begin{equation}
    T + R = 1 - \frac{4 \chi_{1,i}^2}{|f(\chi)|^2} \left\{ \chi^2_r +\chi_{1,i}^2- \left(\sqrt{\frac{\varepsilon_1}{\mu_1}} + \sqrt{\frac{\varepsilon_2}{\mu_2}} \right)^2 \right \} < 1,
\end{equation}
and energy is absorbed in the process.
\item {\bf Case C:}  However, if $\chi_{2,i} >\chi_{1,i} >0$,then $T + R >1$, which suggests some gain and is unphysical for a passive medium.
\end{itemize}

The origin of such contradiction lies in our assumptions regarding the material parameters. As we discuss in the Section~\ref{sec:Conditions}, an imaginary correction to the Tellegen response is necessarily accompanied by the imaginary corrections to the permittivity and permeability of the medium, which allows to restore the passivity.



\section{Reflection and transmission of a plane electromagnetic wave incident at the slab with complex $\chi$.}

\begin{figure}[h!]
  \centering
\includegraphics[width=0.70\textwidth]{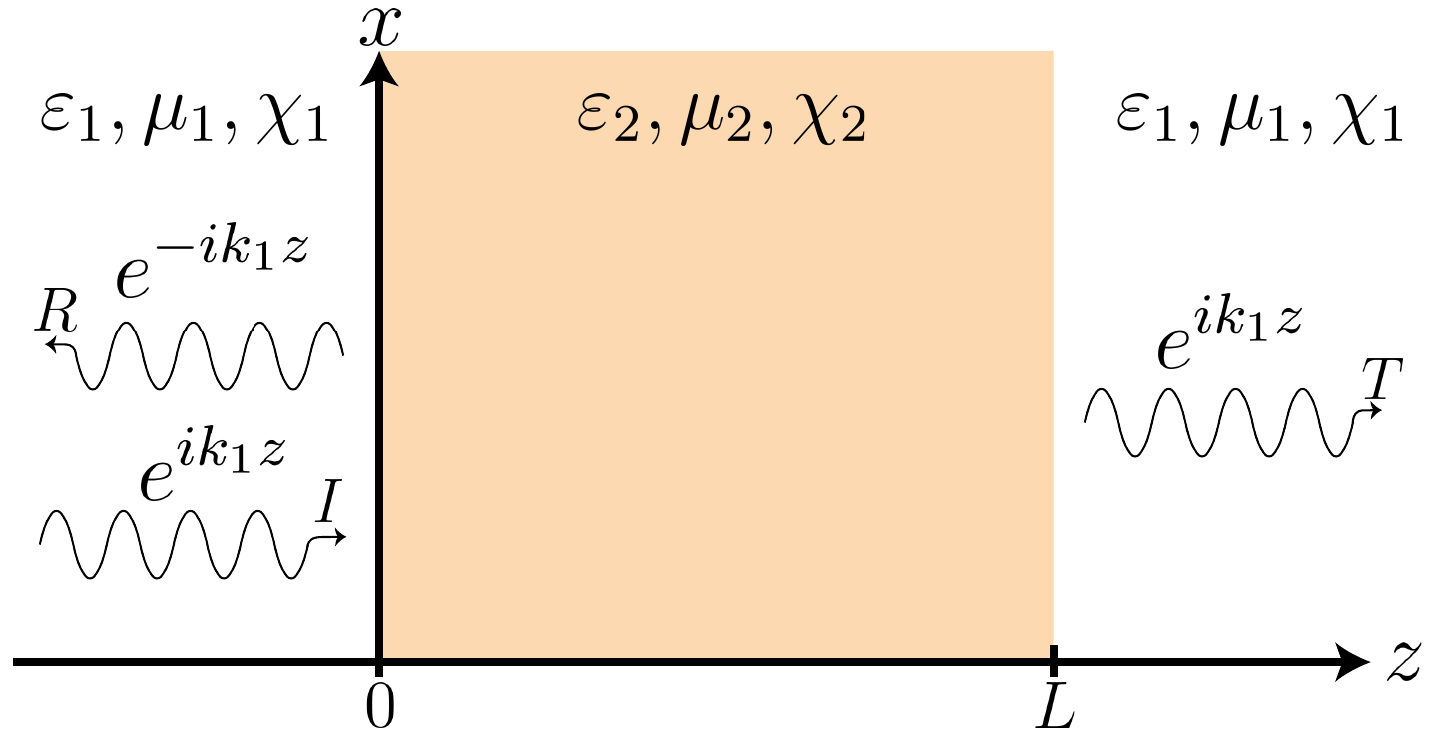}
  \caption{Reflection and transmission of a plane electromagnetic wave through a slab with complex Tellegen response $\chi$.} 
  \label{fig:axionslabcomplex2}
\end{figure}

Next we examine more physical situation when the plane wave is incident at the slab possessing complex axion response, Fig.~\ref{fig:axionslabcomplex2}. Our goal  is to derive the amplitude reflection and transmission matrices.

We apply the same transfer matrix technique using the basis of $\mathbf{E}_t$ and $\mathbf{H}_t$ fields. By considering the propagation inside the complex axion slab of thickness $L$ and Eq. (\ref{transfermatrixeqaxion}), we write the full transfer matrix as

\begin{equation}
    \hat{M} = \hat{M}_{2\rightarrow 1} \hat{M}(L) \hat{M}_{1\rightarrow 2} 
\end{equation}
with

\begin{align}
    \hat{M}_{1\rightarrow 2 } & =  \begin{pmatrix}
        \hphantom{-}1 & \hphantom{-}0 & \hphantom{-}0 & \hphantom{-}0 \\
        \hphantom{-}0 & \hphantom{-}1 & \hphantom{-}0 & \hphantom{-}0 \\
        \hphantom{-}0 & -\chi & \hphantom{-}1 & \hphantom{-}0 \\
        \hphantom{-}\chi & \hphantom{-}0 & \hphantom{-}0 & \hphantom{-}1
    \end{pmatrix}, & 
    \hat{M}_{2\rightarrow 1 } & = \begin{pmatrix}
        \hphantom{-}1 & \hphantom{-}0 & \hphantom{-}0 & \hphantom{-}0 \\
        \hphantom{-}0 & \hphantom{-}1 & \hphantom{-}0 & \hphantom{-}0 \\
        \hphantom{-}0 & \hphantom{-}\chi & \hphantom{-}1 & \hphantom{-}0 \\
        -\chi & \hphantom{-}0 & \hphantom{-}0 & \hphantom{-}1
    \end{pmatrix}
\label{timeboundaryaxiontwoways},\end{align}
and
\begin{equation}
    \hat{M}(L) =
   \begin{pmatrix}
  \vphantom{\sqrt{\dfrac{\varepsilon_1}{\mu_1}}} \cos(k' L) & i \sin(k' L) \sqrt{\dfrac{\mu'}{\varepsilon'}} \\
    i \sin(k' L) \sqrt{\dfrac{\varepsilon'}{\mu'}} & \cos(k' L)
    \end{pmatrix},
\end{equation}
where $\chi = \chi_1-\chi_2$, $k_2 =k_0 \sqrt{\mu_2\varepsilon_2}$, and the boundaries are located at $z=0$ and $z = L$. The components of the reflection and transmission matrices for the slab propagation in the general case are

\begin{align}
    t_{11} \hphantom{-} = \hphantom{-}t_{22} & = \frac{2 i}{g(\chi)} \sqrt{\frac{\varepsilon_1\varepsilon_2}{\mu_1 \mu_2}}, \label{Txx:complexaxionslab}\\
    t_{12} \hphantom{-} = \hphantom{-}t_{21} & = 0, \label{Txy:complexaxionslab}\\
    r_{11} \hphantom{-} = \hphantom{-}r_{22} & =  \frac{1}{g(\chi)} \left( \frac{\varepsilon_1}{\mu_1} - \frac{\varepsilon_2}{\mu_2} - \chi^2\right) \sin(k_2 L), \label{Rxx:complexaxionslab} \\
    r_{12} \hphantom{-} = - r_{21} & = - \frac{2}{g(\chi)}\sqrt{\dfrac{\varepsilon_1}{\mu_1}} \chi \sin(k_2 L),\label{Rxy:complexaxionslab}
\end{align}
where 
\begin{equation}
    g(\chi) = 2i\sqrt{\frac{\varepsilon_1\varepsilon_2}{\mu_1 \mu_2}} \cos(k_2 L) +\left( \frac{\varepsilon_1}{\mu_1} + \frac{\varepsilon_2}{\mu_2} + \chi^2\right) \sin(k_2 L).
\end{equation}

The intensity transmission and reflection coefficients are simply:
\begin{align}
      T & = |t_{11}|^2+|t_{21}|^2, & R & = |r_{11}|^2+|r_{21}|^2.
\end{align}

\begin{figure}[t!]
  \centering
\includegraphics[width=0.95\textwidth]{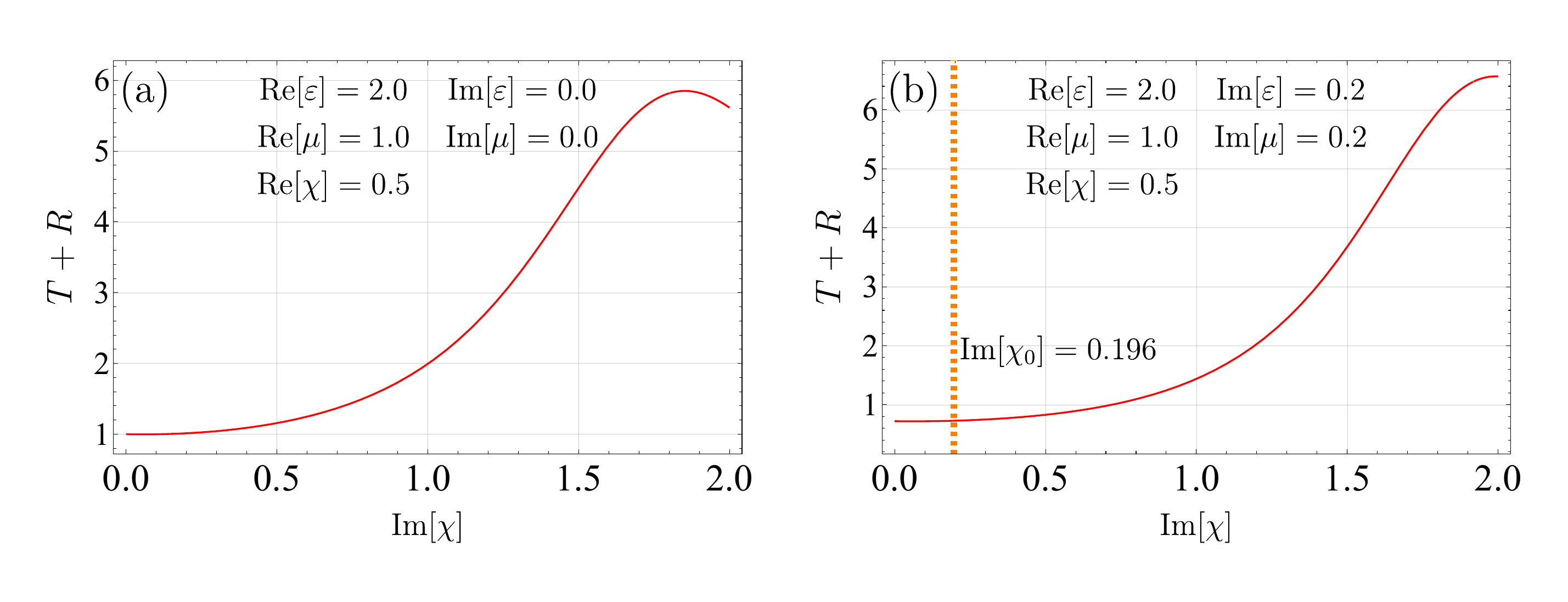}
  \caption{Probing the energy conservation for the light incident at the slab with complex Tellegen response. (a) Introducing imaginary $\chi$ with purely real $\eps$ and $\mu$ results in unphysical behavior $T+R>1$. (b) Dissipation emerges when complex permittivity and permeability are introduced, restoring energy conservation up to a threshold value of  $\chi''$. The threshold is shown by the vertical orange line. Both plots assume $(\omega/c) L = 1$.} 
  \label{fig:testenergy}
\end{figure}

We again probe the energy conservation by plotting the sum $T+R$ in Fig.~\ref{fig:testenergy}. In agreement with the previous section, assumption of complex $\chi$ along with purely real $\eps$ and $\mu$ immediately leads to the inconsistency illustrated in Fig.~\ref{fig:testenergy}(a). However, this artifact is readily removed by incorporating imaginary corrections to $\eps$ and $\mu$, see Fig.~\ref{fig:testenergy}(b). The vertical dashed line shows the maximal attainable value of $\text{Im}\,\chi$ given the prescribed imaginary parts $\eps''$ and $\mu''$. This limitation on $\chi''$ is further discussed in the next section.





\section{Conditions for the effective material parameters}\label{sec:Conditions}

In this section, we derive the constraints on the material parameters for a medium exhibiting complex Tellegen response or complex gyrotropy using the passivity condition. The latter guarantees that the arbitrary configuration of electromagnetic field created in the medium leads to the dissipation of energy.

\subsection{Limitations on complex axion response}

The restriction on the imaginary part of complex Tellegen response is obtained from considerations of the passivity of the medium: any configuration of electromagnetic field passing through this medium should lose energy, but not gain it. For monochromatic waves,
\(\mathbf{E}(t) = \mathbf{E}_0 e^{- i \omega t} + \mathbf{E}_0^* e^{ i \omega t}\), the dissipation power is expressed as
\begin{equation} \label{eq:diss}
    q = -\langle \nabla \cdot \mathbf{S} \rangle = - \langle \frac{c}{4 \pi}  \nabla \cdot (\mathbf{E} \times \mathbf{H})\rangle = \langle \frac{1}{4 \pi} \left( \mathbf{H} \cdot \frac{\partial \mathbf{B}}{\partial t} + \mathbf{E} \cdot \frac{\partial \mathbf{D}}{\partial t} \right) \rangle =  \frac{\omega}{2 \pi} \operatorname{Im}(\mathbf{H}^* \cdot\mathbf{B} + \mathbf{E}^* \cdot \mathbf{D}).
\end{equation}
where $\mathbf{S}$ is the Poynting vector. Substituting the constitutive relations for a complex axion response:
\begin{align}
    \mathbf{D} & = \varepsilon \mathbf{E} + \chi  \mathbf{B}, & \mathbf{H} & = \mu^{-1}\mathbf{B} - \chi  \mathbf{E},
\end{align}
the dissipation power becomes:
\begin{equation} 
\begin{split} 
q &= \frac{c}{2\pi} \omega \operatorname{Im}\left( \mathbf{E}_0^* \cdot \mathbf{D}_0 - \mathbf{B}_0^* \cdot \mathbf{H}_0 \right) \\ & = \frac{c}{2\pi} \omega \operatorname{Im}\left( \varepsilon |\mathbf{E}_0|^2 - \mu^{-1} |\mathbf{B}_0|^2 + \chi \mathbf{E}_0^* \cdot \mathbf{B}_0 - \chi^* \mathbf{E}_0 \cdot \mathbf{B}_0^* \right). \end{split} 
\end{equation}
The dissipation must be nonnegative for any configuration of the field, including the situation $\mathbf{B}_0 \parallel \mathbf{E}_0$. Writing the parameters in terms of real and imaginary parts \(\varepsilon = \varepsilon' + i \varepsilon'' \), \(\mu = \mu' + i \mu'' \) with \( \varepsilon', \varepsilon'', \mu', \mu'' \in \mathbb{R} \) and assuming $\mathbf{B}_0 = a \mathbf{E}_0$, $a \in \mathbb{R}$, we recover the expression for the dissipation
\begin{equation}
    q = \frac{c}{2\pi} \omega \left[ \varepsilon'' |\mathbf{E}_0|^2 +\frac{\mu''}{|\mu|^2}|\mathbf{B}_0|^2 \pm 2 \chi'' |\mathbf{E}_0| |\mathbf{B}_0| \right] \geq 0.
\end{equation}
It is then straightforward to deduce that $\varepsilon''\geq 0$, $\mu''\geq 0$, and 
\begin{equation}\label{eq:restriction}
    |\chi''| \leq \frac{\sqrt{\varepsilon'' \mu''}}{|\mu|}.
\end{equation}

It is important to emphasize that the presence of dissipation in both $\varepsilon$ and $\mu$ is a fundamental requirement for sustaining a nonzero imaginary part of $\chi$. Without dissipation $(\varepsilon''=\mu''=0)$, the imaginary component of $\chi$ is prohibited, as it would violate the passivity condition and lead to unphysical energy gain. Conversely, the real part of $\chi$ remains unrestricted by this condition and can vanish without affecting the constraints on the medium's passivity. These findings are consistent with earlier works, such as Refs.~\cite{lindell1992methods, lindell1994electromagnetic}.

\subsection{Limitations on complex gyrotropy }

Analogously, we derive the restrictions for the imaginary part of gyrotropy. In the case of the material with gyrotropic $\hat{\eps}$  the material equations read \(\mathbf{D} = \hat{\varepsilon}\, \mathbf{E}, \mathbf{H} =\mathbf{B}\), where \(\hat{\varepsilon}\) the dielectric constant tensor is of the form:

\begin{align}
    D_i & = \eps_{ij} \; E_j, &  \eps_{ij} & =  
    \begin{pmatrix}
        \eps & i g &  0\\
        -ig & \eps &0  \\
         0 & 0  &\mu  \\
    \end{pmatrix},
\end{align}
In this case, the expression for the dissipation rate, (\ref{eq:diss}), is
\begin{equation}\label{}
    q = \frac{\omega}{2 \pi} \operatorname{Im}(\mathbf{E}^* \cdot \mathbf{D}) = \frac{\omega}{2\pi} \operatorname{Im}\left( E_i^* E_j \varepsilon_{i j}\right).
\end{equation}
By writing explicitly the parameters \(\varepsilon = \varepsilon' + i \varepsilon'', g = g' + ig'' \), with \( \varepsilon', \varepsilon'', g', g'' \in \Re \), we have
\begin{equation}
    q = \frac{\omega}{8 \pi} \left[ \varepsilon''|E_z|^2 + (|E_x|^2 + |E_y|^2) \varepsilon'' + 2 g'' \text{Im}(E_x E_y^*) \right].
\end{equation}
Given that the dissipation rate is non-negative for any configuration of the electromagnetic field, we can choose the propagation of light along the \(z\) axis. For circularly polarized waves, we can write
\begin{equation} \label{eq_quadr}
    \varepsilon'' (|E_x|^2 + |E_y|^2) + 2 g'' \text{Im}(E_x E_y^*) = \varepsilon'' (|E_x|^2 + |E_y|^2) \pm 2 g'' |E_x||E_y| \geq 0,
\end{equation}
from which we can obtain the inequalities for the material parameters as follows:
\begin{equation}
    \begin{gathered}
        \varepsilon'' \geq 0, \\
        |g''| \leq \varepsilon''.
    \end{gathered}
\end{equation}
These constraints imply that a complex gyrotropy, $g = g' + ig''$, can only exist if the permittivity is also complex and its positive imaginary part $\varepsilon''$ is larger or equal to the imaginary part of  gyrotropy  $|g''|$. 

Within the effective medium description of multilayered structures, this relationship further indicates that the complex Tellegen response requires complex permittivity to satisfy passivity. Importantly, in a metamaterial constructed from the layers with gyrotropic permittivity, no effective correction to permeability $\mu$ emerges. However, spatial dispersion effects in the structure mimic to some extent the permeability-like contributions. This characteristic enables the design of metamaterials that exhibit complex Tellegen response driven primarily by the complex permittivity.

\section{Multilayered gyrotropic medium with effective complex Tellegen response}

In this section, we present numerical simulations of a multilayered axion metamaterial composed of alternating layers with opposite complex gyrotropy, as proposed in the main text. The metamaterial’s behavior is analyzed using the transfer matrix method, and the results are compared with the effective medium description of a slab featuring complex Tellegen response. The Tellegen response and permittivity are described by Eqs. (8) and (9) of the main text.

First, we consider a 1D photonic crystal composed of two gyrotropic layers, $A$ and $B$,  with periodicity given by $\vc{R} = D \vc{\hat{z}}$, where $D=d_A+d_B$ is the size of the unit cell, and $d_\alpha$ is the thickness of the $\alpha$ layer. The permittivity tensor is defined by the expression:

\begin{align}\label{eq:gyroepsilon}
\hat{\varepsilon}(z + D) =\hat{\varepsilon}(z) & = \begin{pmatrix}
 \varepsilon(z) & i g(z) &  0\\
-ig(z) &  \varepsilon(z) &0  \\
 0 & 0  &  \varepsilon(z) \\
\end{pmatrix}, & \hat{\mu}(z+D) = \hat{\mu}(z) & =\mu(z).
\end{align}
We assume $\varepsilon(z)$, $\mu(z)$ are real, and $g(z)$ can be complex. The  constitutive relations for a single layer $\alpha=A,B$ are 

\begin{align} \label{eq:eps_gyr}
    D^i_\alpha & = \eps_\alpha^{ij} \; E_\alpha^j, &  \eps_\alpha^{ij} & =  
    \begin{pmatrix}
        \eps_\alpha & i g_\alpha &  0\\
        -ig_\alpha & \eps_\alpha &0  \\
         0 & 0  &\mu_\alpha  \\
    \end{pmatrix},
\end{align}
where we will consider $g_A = g = -g_B$. Using a chiral basis, $\mathbf{E}_\alpha = E_\alpha \,\hat{\vc{x}}+ i\eta E_\alpha \,\hat{\vc{y}}$, where $\eta = \pm 1$ for right- and left-handed polarization, we can write the $\mathbf{D}$ fields as

\begin{align}
    D_{\alpha,\eta} & = \eps_{\alpha,\eta} E_{\alpha,\eta}, & \eps_{\alpha,\eta} & =  \varepsilon_\alpha-\eta g_\alpha.
\end{align}
Therefore, we can construct the electric field propagating in a layer $\alpha$ as~\cite{Araujo2021Nov}

\begin{equation}
    \mathbf{E}_{\alpha,\eta} = E_{\alpha,\eta}\; e^{i k_{\alpha,\eta} t} \hat{{v}}_\eta + E'_{\alpha,\eta} \; e^{-i k_{\alpha,\eta} t} \hat{{v}}_\eta,
\label{Echiralbasis}\end{equation}
and
\begin{equation}
    \mathbf{H}_{\alpha,\eta} = -i\eta \sqrt{\dfrac{\varepsilon_{\alpha,\eta}}{\mu_{\alpha}}} E_{\alpha,\eta}\; e^{i k_{\alpha,\eta} t} \hat{{v}}_\eta + i\eta \sqrt{\dfrac{\varepsilon_{\alpha,\eta}}{\mu_{\alpha}}} E'_{\alpha,\eta}\; e^{-i k_{\alpha,\eta} t} \hat{{v}}_\eta,
\label{Hchiralbasis}\end{equation}
where 
\begin{equation}
    k_{\alpha,\eta} = k_0 \sqrt{\varepsilon_{\alpha,\eta} \mu_{\alpha}},
\label{dispersionrelationgyrolayer}\end{equation}
is the wavevector with $k_0 = \omega/c$ a constant, $E_{\alpha,\eta}$ and $E'_{\alpha,\eta}$ are the amplitudes of the wave travelling to the right(left). Here, $\hat{v}_\eta= \hat{v}_\pm$ is given by
\begin{equation}
    \hat{v}_\eta = \frac{1}{\sqrt{2}}
    \begin{pmatrix}
        1 \\
        i \eta \\
        0 
    \end{pmatrix}.
\end{equation}
By applying the conventional boundary conditions, i.e., the continuity of the tangential components of the $\vc{H}$ and $\vc{E}$ fields, in Eqs. (\ref{Echiralbasis}),(\ref{Hchiralbasis}) we obtain

\begin{align}
    E_A + E'_A & = E_B + E'_B,\label{boundarygyrotropictime1} \\
    \xi_A E_A -  \xi_A E'_A & =  \xi_B  E_B - \xi_B  E'_B, \label{boundarygyrotropictime2}
\end{align}
where $\xi_\alpha = -i\eta \sqrt{\varepsilon_{\alpha,\eta} / \mu_{\alpha}}$. From this last set of equations, (\ref{boundarygyrotropictime1}),(\ref{boundarygyrotropictime2}) we can write the $2 \times 2$ interface matrix from layer $A$ to $B$:

\begin{align}
    \begin{pmatrix}
        E_B\\
        E'_B
    \end{pmatrix} & = M_{AB} 
    \begin{pmatrix}
        E_A\\
        E'_A
    \end{pmatrix}, & M_{AB} & = \frac{1}{2}
    \begin{pmatrix}
        1 + \dfrac{Y_A}{Y_B} &  1 - \dfrac{Y_A}{Y_B} \\
        1 - \dfrac{Y_A}{Y_B} &  1 + \dfrac{Y_A}{Y_B}
    \end{pmatrix}.
\end{align}
Here, we consider a basis with the amplitude vector like $(E_{\alpha,+},E'_{\alpha,+},E_{\alpha,-},E'_{\alpha,-})^{T}$, where $+(-)$ describes the right- (left-) handed polarization. The $4\times 4$ interface matrix is given by

\begin{equation}
    \begin{pmatrix}
         \vphantom{\dfrac{Y_{A,+}}{Y_{B,+}}} E_{B,+} \\
         \vphantom{\dfrac{Y_{A,+}}{Y_{B,+}}} E'_{B,+} \\
         \vphantom{\dfrac{Y_{A,+}}{Y_{B,+}}} E_{B,-} \\
         \vphantom{\dfrac{Y_{A,+}}{Y_{B,+}}} E'_{B,-}
    \end{pmatrix} = \frac{1}{2}
    \begin{pmatrix}
         1 + \dfrac{Y_{A,+}}{Y_{B,+}} &  1 - \dfrac{Y_{A,+}}{Y_{B,+}} & 0 & 0 \\
         1 - \dfrac{Y_{A,+}}{Y_{B,+}} & 1 + \dfrac{Y_{A,+}}{Y_{B,+}}&  0 & 0 \\
         0 & 0 & 1 + \dfrac{Y_{A,-}}{Y_{B,-}} &  1 - \dfrac{Y_{A,-}}{Y_{B,-}} \\
         0 & 0 & 1 - \dfrac{Y_{A,-}}{Y_{B,-}} &  1 + \dfrac{Y_{A,-}}{Y_{B,-}}
    \end{pmatrix}
    \begin{pmatrix}
         \vphantom{\dfrac{Y_{A,+}}{Y_{B,+}}} E_{A,+} \\
         \vphantom{\dfrac{Y_{A,+}}{Y_{B,+}}} E'_{A,+} \\
         \vphantom{\dfrac{Y_{A,+}}{Y_{B,+}}} E_{A,-} \\
         \vphantom{\dfrac{Y_{A,+}}{Y_{B,+}}} E'_{A,-}
    \end{pmatrix},
\end{equation}
where we define $Y_{\alpha, \eta} = \sqrt{\varepsilon_{\alpha,\eta} / \mu_{\alpha}}$. Finally, for the case of an electromagnetic wave propagating inside the medium $\alpha=A,B$, with thickness $d_\alpha$ and wavevector $k_{\alpha,\eta}$, the $4\times 4$ bulk-propagation matrix is given by

\begin{equation}
    M_\alpha = M_\alpha(d_\alpha) = 
    \begin{pmatrix}
        e^{i k_{\alpha,+} d_\alpha} & 0 & 0 & 0 \\
        0 & e^{-i k_{\alpha,+} d_\alpha} & 0 & 0 \\
        0 & 0 & e^{i k_{\alpha,-} d_\alpha} & 0 \\
        0 & 0 & 0 & e^{-i k_{\alpha,-} d_\alpha}
    \end{pmatrix}.
\end{equation}
The resultant transfer matrix describing the gyrotropic multilayered  structure with even number of $2N$ layers then reads

\begin{equation}
    M_{\text{slab}} = M_{B-\text{air}} M_{AB} (M_{BA} M_{B} M_{AB} M_{A})^{2N} M_{\text{air}-A}  = 
    \begin{pmatrix}
        M^+_{\text{slab}} & 0 \\
        0 & M^-_{\text{slab}}
    \end{pmatrix}.
\end{equation}
This matrix, describes the following system,

\begin{align}
    \begin{pmatrix}
        E^t_\eta \\
        0
    \end{pmatrix} & = M^\eta_{\text{slab}} 
    \begin{pmatrix}
        E^i_\eta \\
        E^r_\eta
    \end{pmatrix}, & M^\eta_{\text{slab}} & =
    \begin{pmatrix}
        M^\eta_{11} & M^\eta_{12} \\
        M^\eta_{21} & M^\eta_{22}
    \end{pmatrix}.
\end{align}
From which we can deduce the amplitude of the transmitted and reflected wave as for both polarizations 

\begin{align}
    E^t_\eta & =\frac{\text{det}(M^\eta)}{M^\eta_{22}} E^i_\eta &  E^r_\eta & = -\frac{M^\eta_{21}}{M^\eta_{22}} E^i_\eta.
\end{align}
Assuming that the incident wave is polarized along the $x$ axis, we write the transmitted and reflected fields as

\begin{eqnarray}
    \mathbf{E}^t & = & \frac{1}{2}\left(\frac{\text{det}(M^+)}{M^+_{22}} + \frac{\text{det}(M^-)}{M^-_{22}} \right) E^i \, \hat{\vc{x}} + \frac{i}{2}\left(\frac{\text{det}(M^+)}{M^+_{22}} - \frac{\text{det}(M^-)}{M^-_{22}} \right) E^i \, \hat{\vc{y}}, \\
    \mathbf{E}^r & = & - \frac{1}{2}\left( \frac{M^+_{21}}{M^+_{22}} + \frac{M^-_{21}}{M^-_{22}} \right) E^i \, \hat{\vc{x}} - \frac{i}{2}\left( \frac{M^+_{21}}{M^+_{22}} - \frac{M^-_{21}}{M^-_{22}} \right) E^i \, \hat{\vc{y}}.
\end{eqnarray}
which yields the co- and cross-polarized transmission and reflection in terms of the components of the transfer-matrix for the full multilayered slab
\begin{align}
    T_{xx} & = \frac{1}{2}\left(\frac{\text{det}(M^+)}{M^+_{22}} + \frac{\text{det}(M^-)}{M^-_{22}} \right), & T_{xy} & = \frac{i}{2}\left(\frac{\text{det}(M^+)}{M^+_{22}} - \frac{\text{det}(M^-)}{M^-_{22}} \right), \label{T:photoniccrystal}\\
    R_{xx} & =- \frac{1}{2}\left( \frac{M^+_{21}}{M^+_{22}} + \frac{M^-_{21}}{M^-_{22}} \right), & R_{xy} & = - \frac{i}{2}\left( \frac{M^+_{21}}{M^+_{22}} -\frac{M^-_{21}}{M^-_{22}} \right).\label{R:photoniccrystal}
\end{align}
We validate the effective medium model by simulating a multilayered structure with 50 unit cells, each consisting of two layers of equal thickness  $a/2$ and opposite gyrotropy $\pm g_0$, $ g_0 = 0.5 + i 0.01$. For simplicity, we assume that the permittivity and permeability of the gyrotropic layers are equal to 1. The transmission and reflection coefficients of the photonic structure are computed using Eqs~(\ref{T:photoniccrystal}),(\ref{R:photoniccrystal}). 

The effective axion response depends on the way the structure is terminated. If the gyrotropy of the first layer is positive, the real part of the effective axion response is also positive and an effective material paramters are given by:  
\begin{align}
    \chi_\text{eff} & =  \frac{\pi}{2} \, g_0 \,\xi, & \eps_\text{eff} & =  \eps \left( 1-\frac{\pi^2}{12}\frac{g_0^2}{\eps\mu} \xi^2\right),
\label{effectiveaxionfieldandmu}\end{align}
where $g_0$ is the gyrotropy of the first layer of the metamaterial, $\xi = a/\lambda$ is the period-to-wavelength ratio, and $\lambda$ is the vacuum wavelength. The permeability remains unchanged, $\mu_\text{eff} = \mu$. The transmission and reflection coefficients for an axion slab have the analytical forms given by Eqs.~(\ref{Txx:complexaxionslab})-(\ref{Rxy:complexaxionslab}). For the permittivity and permeability, we set $\varepsilon_1  =\mu_1 = \mu_2 = 1$, and $\varepsilon_2 = \eps_\text{eff}$ to model the complex axion slab in vacuum. The phase accumulation inside the slab is given by:
\begin{equation}
 k_2 L = 2\pi\sqrt{\mu_{\text{eff}} \, \varepsilon_{\text{eff}}} N \xi,
\end{equation}
where $N = 50$ represents the number of periods in the multilayered structure. 

The thickness of the slab is fixed at $L = N a $. By varying the wavelength $\lambda$, we  plot the reflection and transmission coefficients both for our photonic structure and fpr the effective medium model in terms of the period-to-wavelength ratio $\xi$.

\begin{figure}[h!]
\center{\includegraphics[width=0.8\linewidth]{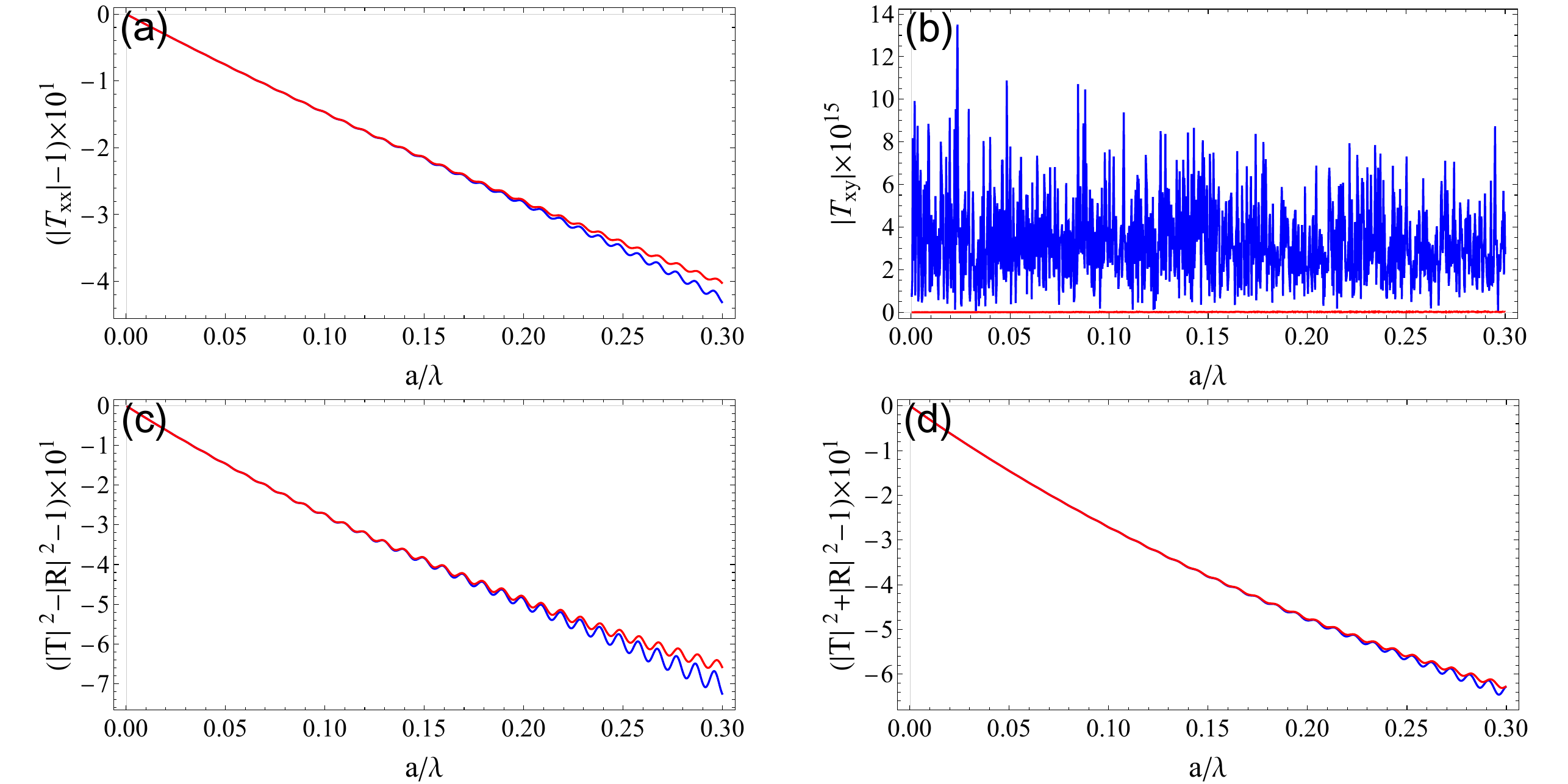}}
\caption{Validating complex effective Tellegen response. Blue lines correspond to the transfer-matrix method applied to the multilayered structure, and red lines to the effective medium model. Panels (a,b) correspond to the co- and cross-polarized transmission, respectively, panel (c) displays the
momentum non-conservation $|T|^2 - |R|^2 \neq 1$, and panel (d) shows energy non-conservation $|T|^2 + |R|^2 \neq 1$. An arrangement of 50 layers with the gyrotropic permittivity is considered.}
\label{validacionsuple}
\end{figure}

In the main text [Fig. 3], we validate the effective axion response. It is evident that for subwavelength modulation with $\xi \ll 1$, the transmission and reflection coefficients of the multilayered structure closely match those of the effective medium. As $\xi$ increases, the effective medium solution starts to deviate from the real multilayered structure. In Fig. \ref{validacionsuple}(a-d) we present the results complementing those in the main text. We observe that the co-polarized transmission is quite close to unity and monotonically decreases with $\xi$ experiencing the oscillations  [Fig.~\ref{validacionsuple}(a)]. The cross-polarized transmission vanishes [Fig.~\ref{validacionsuple}(b)], while the energy carried by the wave is not conserved, i.e. $|T|^2+|R|^2<1$ [Fig.~~\ref{validacionsuple}(d)]. The momentum of the wave quantified by $|T|^2-|R|^2$ is also not conserved [Fig.~~\ref{validacionsuple}(c)] since the spatial boundaries break the translational invariance of the system.

\section{Retrieval of the material properties via Stokes parameters}

Material parameters are typically determined from the measurements of the reflection and transmission coefficients using the celebrated Nicolson-Ross-Weir method \cite{nicolson1970, weir1974}.


In this section, we further advance this technique to show how real and imaginary parts of the Tellegen response could be retrieved. For that purpose, one needs to assess not only the intensity, but also the polarization of the reflected light characterized by the set of Stokes parameters:


\begin{equation}\label{eq:stokes}
    \begin{gathered}
        I = |E_x|^2 + |E_y|^2, \\
        Q = |E_x|^2 - |E_y|^2, \\
        U = 2 \ \text{Re} \left[E_x E_y^*\right], \\
        V = -2  \text{Im} \left[E_x E_y^*\right],
    \end{gathered}
\end{equation}
where \( I \) represents the total intensity, while \( Q, U, V \) describe the polarization state. These parameters are related through the Poincaré sphere identity:
\begin{equation} 
I^2 = Q^2 + U^2 + V^2. 
\end{equation}
Since only three of these parameters are independent, an additional parameter is required to fully determine the four unknowns \( \varepsilon' \), \( \varepsilon'' \), \( \chi' \), and \( \chi'' \).  We assume that the transmitted light intensity,  $I_T$,  is also known, which allows us to solve for all four unknowns.


The relationship between the reflected and incident electric fields is given by the reflection matrix:
\begin{equation}
    \begin{pmatrix}
        E_x^r \\
        E_y^r
    \end{pmatrix} =
    \begin{pmatrix}
        R_{xx} & R_{xy} \\
        R_{yx} & R_{yy}
    \end{pmatrix}
    \begin{pmatrix}
        E_x^i \\
        E_y^i
    \end{pmatrix},
\end{equation}
where the reflection coefficients for a complex axion slab with permittivity $\varepsilon = \varepsilon' + i \varepsilon''$ and axion coupling $\chi = \chi' + \chi''$, with $\varepsilon',\varepsilon'',\chi', \chi'' \in \mathbb{R}$, are given by Eqs.~(\ref{Rxx:complexaxionslab},\ref{Rxy:complexaxionslab}):
\begin{eqnarray}
      R_{xx} \hphantom{-} = \hphantom{-}R_{yy} & = & \dfrac{1}{g(\chi)} \left( 1 - \varepsilon - \chi^2\right) \sin(k L), \label{Rxx:asb} \\
    R_{xy} \hphantom{-} = - R_{yx} & = &  \dfrac{2}{g(\chi)} \chi \sin(k  L),\label{Rxyasab}
\end{eqnarray}
where $k= \omega/c \sqrt{\varepsilon}$,  and the denominator function is
\begin{equation}
    g(\chi) = 2i\sqrt{\varepsilon} \cos(k L) +\left( 1 + \varepsilon + \chi^2\right) \sin(k L).
\end{equation}
Assuming normal incidence with an $x$-polarized plane wave ($E_y^i=0$), the reflected field components are:
\begin{align}
    E_x^r & = R_{xx}  E_x^i, & E_y^r & = R_{yx} E_x^i.
\end{align}
%

The Stokes parameters of the reflected wave are expressed as:

\begin{align}
    I & = \frac{1}{|g(\chi)|^2} \Bigg[|1-\varepsilon -\chi^2|^2 + 4  |\chi|^2 \Bigg] |\sin(kL)|^2 , \\
    Q & = \frac{1}{|g(\chi)|^2} \Bigg[|1-\varepsilon -\chi^2|^2 - 4  |\chi|^2  \Bigg]|\sin(kL)|^2 , \\
    U & = -\frac{4}{|g(\chi)|^2} \text{Re} \Bigg[(1 - \varepsilon -\chi^2)  \chi^* \Bigg]|\sin(kL)|^2 , \\
    V & = \frac{4}{|g(\chi)|^2} \text{Im} \Bigg[(1 - \varepsilon -\chi^2) \chi^* \Bigg] |\sin(kL)|^2 ,
\end{align}
where we assume  $|E_x^i|^2 =1$. Since the system is nonlinear and complex, solving for $\varepsilon', \varepsilon'',\chi',\chi''$  directly from these equations is not straightforward. However, realistically, the Tellegen response $\chi$ is quite weak. We therefore keep only first power of $\chi$ neglecting the terms of the order of $\chi^2$.

\begin{align}
    I_R & \approx \frac{1}{|g(\varepsilon',\varepsilon'')|^2} \Bigg[(1-\varepsilon')^2+\varepsilon''^2 \Bigg] |\sin(kL)|^2,  \\
    Q_R & \approx \frac{1}{|g(\varepsilon',\varepsilon'')|^2} \Bigg[(1-\varepsilon')^2+\varepsilon''^2 \Bigg] |\sin(kL)|^2,  \\
    U_R & \approx -\frac{4}{|g(\varepsilon',\varepsilon'')|^2} \Bigg[(1 - \varepsilon' )\chi' -\varepsilon'' \chi''   \Bigg]|\sin(kL)|^2,  \\
    V_R & \approx -\frac{4}{|g(\varepsilon',\varepsilon'')|^2}  \Bigg[(1 - \varepsilon') \chi'' + \varepsilon'' \chi' \Bigg] |\sin(kL)|^2, 
\end{align}
where
\begin{equation}
    g(\chi) \approx g(\varepsilon',\varepsilon'') = 2i\sqrt{\varepsilon} \cos(k L) +\left( 1 + \varepsilon\right) \sin(k L),
\end{equation}
is independent of $\chi$. As we mentioned earlier, the total intensity of the transmitted wave is required to fully determine the parameters.
\begin{equation}
    I_T \approx \frac{4 \sqrt{\varepsilon'^2+\varepsilon''^2}}{|g(\varepsilon',\varepsilon'')|^2}.  
\end{equation}
If the losses are small, i.e. $\eps''\ll|\eps'|$, we can approximate the sinusoidal functions as
\begin{align}
    \sin(kL) & \approx \sin(k_0 L \sqrt{\varepsilon'}) + i k_0 L \frac{\varepsilon''}{2 \sqrt{\varepsilon'}}\cos(k_0 L \sqrt{\varepsilon'}), & \cos(kL) & \approx \cos(k_0 L \sqrt{\varepsilon'}) - i k_0 L \frac{\varepsilon''}{2 \sqrt{\varepsilon'}}\sin(k_0 L \sqrt{\varepsilon'}), 
\end{align}
where we use

\begin{equation}
    \sqrt{\varepsilon} \approx \sqrt{\varepsilon'} +i\frac{\varepsilon''}{2\sqrt{\varepsilon'}}.
\end{equation} 
\subsection{Retrieval of real and imaginary parts of permittivity}

The real and imaginary part of permittivity can be recovered using the standard Nicolson-Ross-Weir technique.

In this spirit, we calculate the total intensity, as the sum of the intensity of the reflected and transmitted waves
\begin{equation}
    I = I_T + I_R \approx \frac{1}{|g(\varepsilon',\varepsilon'')|^2} \Bigg[(1-\varepsilon')^2\Bigg] \sin^2(k_0 L \sqrt{\varepsilon'}) + \frac{4 \varepsilon'}{|g(\varepsilon',\varepsilon'')|^2} = \frac{(1+\varepsilon')^2 \sin^2(k_0 L \sqrt{\varepsilon'}) + 4\varepsilon'\cos^2(k_0 L \sqrt{\varepsilon'})}{|g(\varepsilon',\varepsilon'')|^2},
\end{equation}
where
\begin{equation}
    |g(\varepsilon',\varepsilon'')|^2 \approx (1+\varepsilon')^2 \sin^2(k_0 L \sqrt{\varepsilon'}) + 4\varepsilon'\cos^2(k_0 L \sqrt{\varepsilon'}) + \frac{\varepsilon''}{\sqrt{\varepsilon'}} \Big[\sqrt{\varepsilon'}(1+\varepsilon')2 k_0 L + (\varepsilon'-1) \sin(2k_0 L \sqrt{\varepsilon'}) \Big].
\end{equation}
From which we can deduce rather easily the imaginary part of permittivity:
\begin{equation}
    \varepsilon'' = \frac{A}{B} \frac{(1-I)}{I},
\label{Eqimapermitivttiy}\end{equation}
where
\begin{align}
   A & =  (1+\varepsilon')^2 \sin^2(k_0 L \sqrt{\varepsilon'}) + 4\varepsilon'\cos^2(k_0 L \sqrt{\varepsilon'}), & B & = \frac{\sqrt{\varepsilon'}(1+\varepsilon')2 k_0 L + (\varepsilon'-1) \sin(2k_0 L \sqrt{\varepsilon'})}{\sqrt{\varepsilon'}}. 
\end{align}
From the intensity of reflected light we deduce
\begin{equation}
    I_R = \frac{(1-\varepsilon')^2 \sin^2(k_0 L \sqrt{\varepsilon'})}{A + \varepsilon'' B},
\end{equation}
after some manipulation we obtain the following equation:
\begin{equation}
    (1-\varepsilon')^2(1-\alpha_R)\sin^2(k_0 L \sqrt{\varepsilon'}) + 4\varepsilon' = 0,
\end{equation}
where $\alpha_R = I/I_R$. This equation cannot be solved analytically, so we have to resort to numerical methods to obtain a solution. Once $\varepsilon'$ is recovered numerically, we can substitute it back into Eq.~(\ref{Eqimapermitivttiy}) to obtain $\varepsilon''$.

\subsection{Retrieval of real and imaginary parts of Tellegen response}

To assess the Tellegen response, one needs to examine the cross-polarized reflection and analyze the Stokes parameters of the reflected light. We assume that complex permittivity $\eps$ is already recovered. Using the Stokes parameters $U_R$ and $V_R$
\begin{align}
    \frac{U_R}{I_R} & = u_R = - 4 \frac{(1-\varepsilon')\chi' - \varepsilon'' \chi''}{(1-\varepsilon')^2}, &  \frac{V_R}{I_R} & =   v_R = - 4 \frac{(1-\varepsilon')\chi'' + \varepsilon'' \chi'}{(1-\varepsilon')^2},
\end{align}
we extract the Tellegen response as
\begin{align}
    \chi' & = - \frac{(1-\varepsilon')u_R + \varepsilon'' v_R}{4}, & \chi'' & =  -\frac{(1-\varepsilon') v_R -\varepsilon'' u_R}{4}.
\end{align}
To finalize we sum up the results as
\begin{align}
      & (1-\varepsilon')^2(1-\alpha_R)\sin^2(k_0 L \sqrt{\varepsilon'}) + 4\varepsilon' = 0, \\
    & \varepsilon''  = \sqrt{\varepsilon'}\frac{(1+\varepsilon')^2 \sin^2(k_0 L \sqrt{\varepsilon'}) + 4\varepsilon'\cos^2(k_0 L \sqrt{\varepsilon'})}{\sqrt{\varepsilon'}(1+\varepsilon')2 k_0 L + (\varepsilon'-1) \sin(2k_0 L \sqrt{\varepsilon'})} \Bigg[ \frac{1-I}{I} \Bigg], \\
    &\chi'  = - \frac{(1-\varepsilon')u_R + \varepsilon'' v_R}{4}, \label{eq:Rechi}\\
    & \chi''  = -\frac{(1-\varepsilon') v_R -\varepsilon'' u_R}{4},\label{eq:Imchi}
\end{align}
where 
\begin{align}
    \alpha_R & = \frac{I}{I_R}, & v_R & = \frac{V_R}{I_R}, &   u_R & = \frac{U_R}{I_R}.
\end{align}

Equations~\eqref{eq:Rechi},\eqref{eq:Imchi} allow to independently recover both real and imaginary parts of the Tellegen response provided the Stokes parameters of the reflected light are measured experimentally. Therefore, complex Tellegen response can be readily observed providing an effective medium description of complex non-reciprocal photonic structures.


